\documentclass{fmj2010}
\usepackage{graphicx}
\usepackage{natbib}
\usepackage{url}
\def\fgrst{\textit{Fermi Gamma-Ray Space Telescope}}
\def\fermilat{\textit{Fermi}/LAT}
\def\fermi{\textit{Fermi}}

\begin{document}
   \title{The $\gamma$-ray properties of radio-selected extragalactic jets}

   \author{M.~B\"ock\inst{1}
          \and
          M.~Kadler\inst{1,2,3}
	  \and
	  G.~Tosti\inst{4,5}
	  \and
	  T.~H.~Burnett\inst{6}
	  \and
	  C.~M\"uller\inst{1}
	  \and
	  R.~Ojha\inst{7,8} on behalf of the LAT collaboration
	  \and
	  J.~Wilms\inst{1}
	  }
	
   \institute{
     Dr.~Karl Remeis-Observatory \& ECAP, Universit\"at
     Erlangen-N\"urnberg, Sternwartstr.~7, 96049 Bamberg, Germany
     \and
     CRESST/NASA Goddard Space Flight Center, Greenbelt, MD 20771, USA 
     \and
     Universities Space Research Association, 10211 Wincopin Circle, Suite
     500 Columbia, MD 21044, USA      
     \and
     Istituto Nazionale di Fisica Nucleare, Sezione di Perugia, I-06123
     Perugia, Italy
     \and
     Dipartimento di Fisica, Universit\`a degli Studi di Perugia, I-06123
     Perugia, Italy 
     \and
     Department of Physics, University of Washington, Seattle, WA
     98195-1560, USA
     \and
     U.S. Naval Observatory, 3450 Massachusetts Ave. NW, Washington DC
     20392, USA
     \and
     NVI, Inc., 7257 Hanover Parkway, Greenbelt, MD 20770, USA 
     }

   \abstract{ Most extragalactic jets in radio-loud Active Galactic
     Nuclei are bright and variable $\gamma$-ray sources, which are
     continuously monitored with \fermilat. We present the $\gamma$-ray
     properties of the MOJAVE and TANAMI AGN samples of radio-loud AGN.
     Both programs provide properties of the parsec-scale radio jets
     using Very Long Baseline Interferometry (VLBI) techniques. This
     information is important to understand the broad-band emission
     mechanism of these sources. In this work we compare the radio and
     $\gamma$-ray properties of the two samples and present upper limits
     on the $\gamma$-ray flux of the radio-brightest jet sources not yet
     detected by \fermilat.}

   \maketitle

\section{Introduction}

Most extragalactic jets in radio-loud Active Galactic Nuclei (AGN) are
bright and variable $\gamma$-ray sources. The high luminosity and
variability of blazars is explained by a jet orientated close to the
line of sight, in which charged particles are moving with relativistic
speed, thus the emission is Doppler boosted
\citep{Blandford1978,Maraschi1992}. This model is supported by the
apparent superluminal motion which is typically found in the inner
radio-jets of blazars \citep[, and therein]{Lister2009_kinematics}.

Since 2008 August 11, the sky is monitored at $\gamma$-ray energies in
the range from $\sim$20\,MeV to $>$300\,GeV with the Large Area
Telescope (LAT), which is a pair conversion detector on board the
\fgrst{} \citep{Atwood2009}. \fermilat{} scans the whole sky every three
hours and its sky-survey mode is optimized to obtain a uniform exposure
on longer time scales ($>$\,weeks). A large number of AGN were detected
and monitored with \fermilat{} \citep[709 AGN during the first year of
operation,][]{1LAC}.

In this work, we present the $\gamma$-ray properties of two samples from
the two largest ongoing Very Long Baseline Interferometry (VLBI) AGN
monitoring survey programs: Tracking Active Galactic Nuclei with Austral
Milliarcsecond Interferometry \citep[TANAMI;][]{2009Ojha_prelim} and
Monitoring Of Jets in Active galactic nuclei with VLBA Experiments
\citep[MOJAVE;][]{Lister2009_mojave}. The two samples are described in
Sec.~\ref{sec:sample}. For our analysis we use data obtained during the
first year of \fermi{} operations. 

The combined analysis of radio and $\gamma$-ray properties is still in
progress. The detailed presentation of the upper limit analysis, a
discussion of tentative new detections and individual sources is in work
and will be published in an upcoming paper (Abdo et al., 2010, in
prep.).

\section{Sample} \label{sec:sample} In our study, we investigate the
$\gamma$-ray properties of two samples of extragalactic jets: the
MOJAVE\,1 and the TANAMI samples. MOJAVE\,1 is a purely radio-selected
sample of all AGN at declinations $\delta >-20^\circ$, whose 15\,GHz
radio flux density has exceeded 1.5\,Jy (2\,Jy for sources with $\delta
< 0^\circ$) at any epoch between 1994 and 2004
\citep{Lister2009_mojave}. In total, the MOJAVE\,1 sample is comprised
of 135 AGN: 101 quasars, 22 BL\,Lac objects, 8 galaxies, and 4
unclassified objects. About two thirds of the sources in the MOJAVE\,1
sample have been monitored since 1994 as part of the VLBA 2\,cm Survey
\citep{Kellermann1998} with the Very Long Baseline Array
\citep[VLBA;][]{Napier1994}. Since 2002, the MOJAVE\,1 sample has been
observed with the VLBA as part of the MOJAVE
program\footnote{\url{http://www.physics.purdue.edu/astro/MOJAVE/}} (see
also Lister et al.,these proc.).

TANAMI is a radio VLBI monitoring program of currently 75 extragalactic
jets south of $-30^\circ$ declination. The observations are performed
approximately every two months with the telescopes of the Australian
Long Baseline Array \citep[LBA;][]{Norris1988} in combination with
telescopes in South Africa, Antarctica and Chile. A detailed discussion
of the TANAMI program including a description of the initial source
sample, explanations of the observation and data reduction procedures,
and brief notes on the individual sources are given by
\citet{2009Ojha_prelim}. TANAMI contains a radio-selected and a
$\gamma$-ray selected sub-sample \citep[see][for
details]{2009Ojha_prelim}. The radio selected sub-sample supplements the
MOJAVE\,1 sample south of $-30^\circ$ declination. TANAMI observations
are made at 8.4\,GHz and 22\,GHz yielding spectral indices of
parsec-scale jet features (M\"uller er al., these proc.).

\section{Analysis}

To find associations of the AGN in the MOJAVE\,1 and TANAMI sample with
\fermilat{}-detected $\gamma$-ray sources we performed a
positional-association analysis. For each radio position of the sources
in the sample we selected the closest sources in the ``Fermi Large Area
Telescope First Source Catalog'' \citep{1fgl} inside the position
uncertainty region of the $\gamma$-ray source. This approach yielded
unique results for the majority of sources. A comparison with the
results of the automatic source association pipeline of the LAT team
\citep{LBAS2009}, in which the class and physical expectations of the
association is taken into account, confirmed these results.

We used a maximum likelihood analysis \citep{Cash1979,Mattox1996} to 
analyze those AGN which could not be associated with $\gamma$-ray
catalog sources. For the calculation of upper limits on the $\gamma$-ray
flux and the test statistic (TS) measuring the significance we used data
in the range 100\,MeV to 100\,GeV obtained in the first year of
\fermilat{}.

We used the standard \fermilat{} {\it ScienceTools} and the methods
described by \citet{Boeck2009_fs} to obtain the flux, photon index, and
test statistic of each unassociated source. For sources with
$\textnormal{TS}<25$ we calculated upper limits on the flux. Sources
with $\textnormal{TS}\geq25$ were considered as tentative new detections
and in this case we modeled the flux and the spectral index. For the
sources with $\textnormal{TS}<1$ the upper limits can be underestimated,
thus we do not consider these sources in this work. Results for a
Bayesian approach following \citet{1fgl} will be presented in a
following paper.

\section{Results}

\begin{table}
    \label{tab:det} \caption{Total number $N$ of the MOJAVE\,1 and
      TANAMI sources and the number of those detected with \fermilat{}.}
    \centering
    \begin{tabular}{lllll}
	& \multicolumn{2}{l}{MOJAVE\,1} & \multicolumn{2}{l}{TANAMI}\\
	\hline Class & $N$
	&detected & $N$ &
	detected\\	
	\hline
	\hline
	Quasar &        101 &   61 & 32 &    24\\
	BL Lac &         22 &   19 & 10 &    10\\
	Galaxy &          8 &    3 & 11 &     2\\
	Unclassified &    4 &    2 & 22 &    17\\
	\hline
    \end{tabular}
\end{table}

Table~\ref{tab:det} shows the number of quasars, BL\,Lacs, and galaxies
in the MOJAVE\,1 and TANAMI sample which are associated with
$\gamma$-ray sources in the \fermilat{} 1-year source list \citep{1fgl}.
85 out of the 135 AGN in the MOJAVE\,1 sample are detected with
\fermilat{} during its first year of operation. With a detection rate of
$19/22$ most BL\,Lac objects in the sample are detected. $61/101$
quasars are detected and $2/4$ unclassified objects. At $3/8$, the
detection rate of the galaxies is the lowest. In the TANAMI sample 24
out of 32 quasars are detected with \fermilat{}. The detection rates of
unclassified sources is substantially higher in the TANAMI sample than
in the MOJAVE\,1 sample. This is most likely due to the inclusion of a
$\gamma$-ray selected sub-sample and the addition of sources detected by
\fermilat{} after three months \citep{LBAS2009}. The detection rates for
the radio-selected TANAMI sub-sample consisting of 21 AGN are: $9/13$
Quasars, $3/3$ BL\,Lac objects and $2/5$ Galaxies, which is comparable
to the MOJAVE\,1 sample.

\begin{figure*}
    \includegraphics[width=.98\columnwidth]{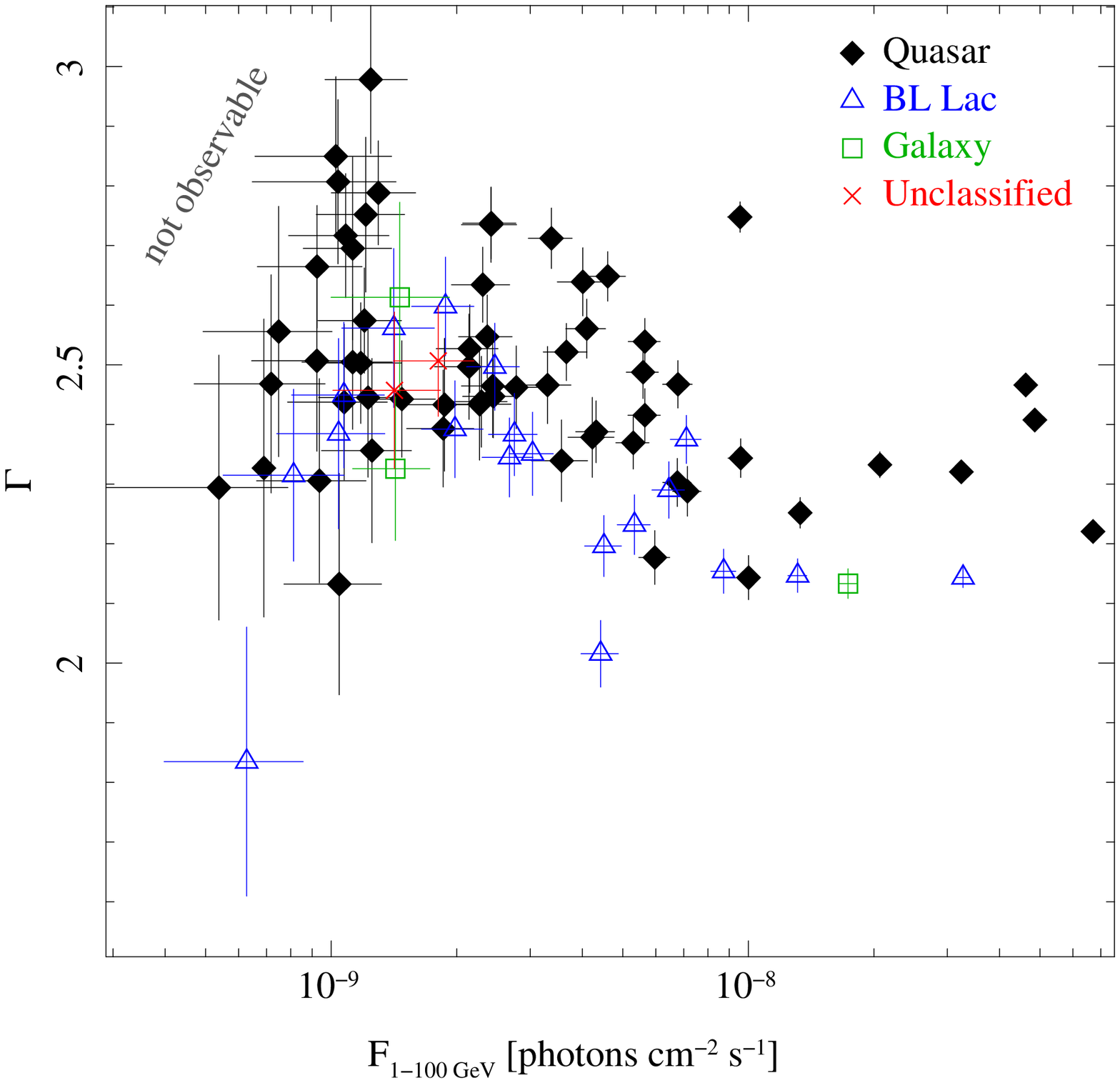}
   \hfill\includegraphics[width=.98\columnwidth]{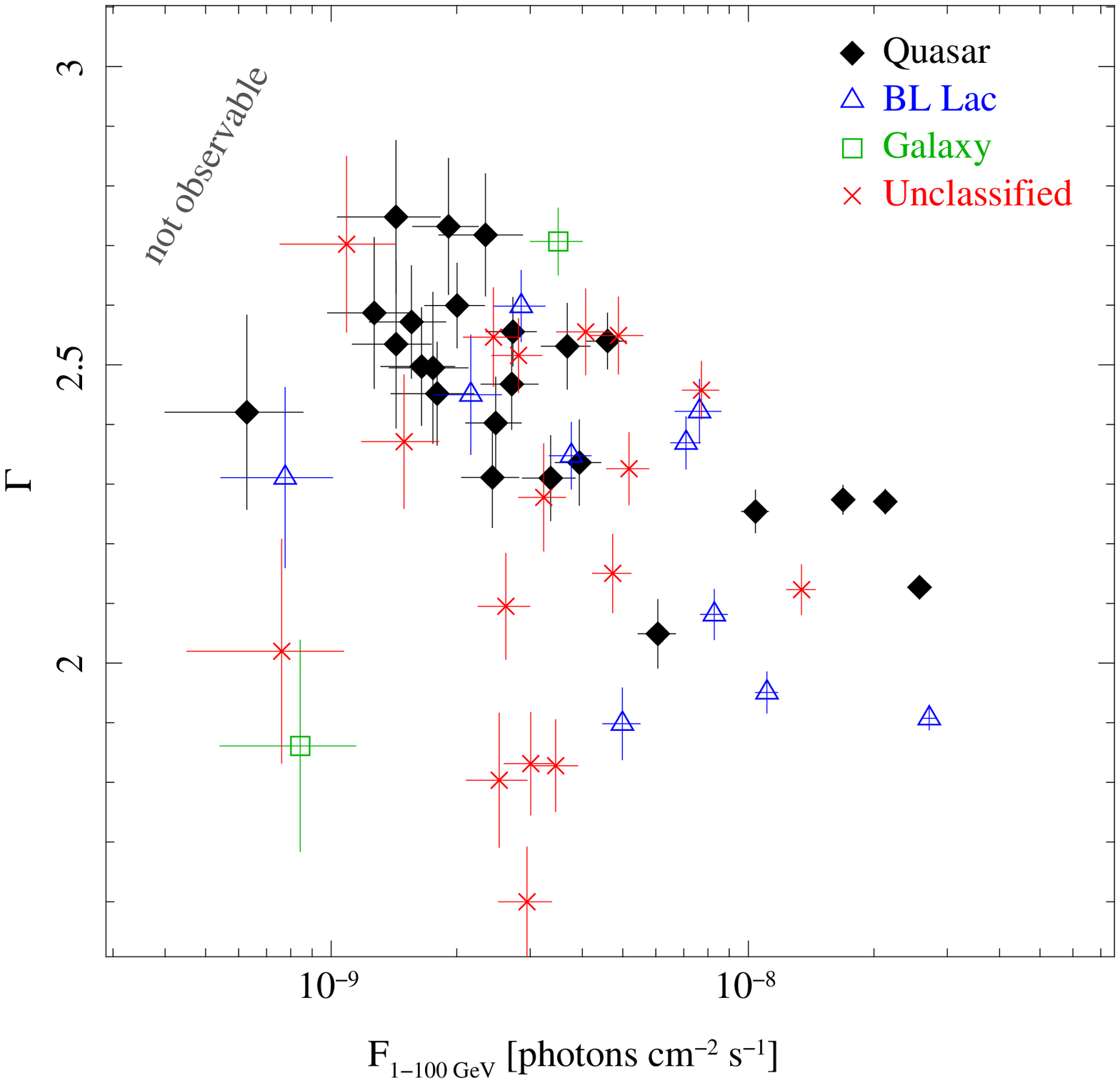}
   \caption{Distribution of flux and photon spectral indices of the
   associated sources in the MOJAVE\,1 (left) and TANAMI (right)
   samples. The flux is given in units of photons\,cm$^{-2}$\,s$^{-1}$
   in the energy range 1--100\,GeV. In the top left corners \fermilat{}
   cannot detect sources.} \label{fig:flux_spec}
\end{figure*}

For the analysis of the associated sources we used the photon spectral
indices from the \fermilat{} first source catalog calculated in the
range of 0.1--100\,GeV, since this value can be obtained even for
relatively weak sources \citep{1LAC}. Figure~\ref{fig:flux_spec} shows
the distribution of flux and photon spectral index averaged over 11
months. In addition to the catalog values we present the values we
obtained for the tentative new detections. 

\begin{figure*}
    \includegraphics[width=\columnwidth]{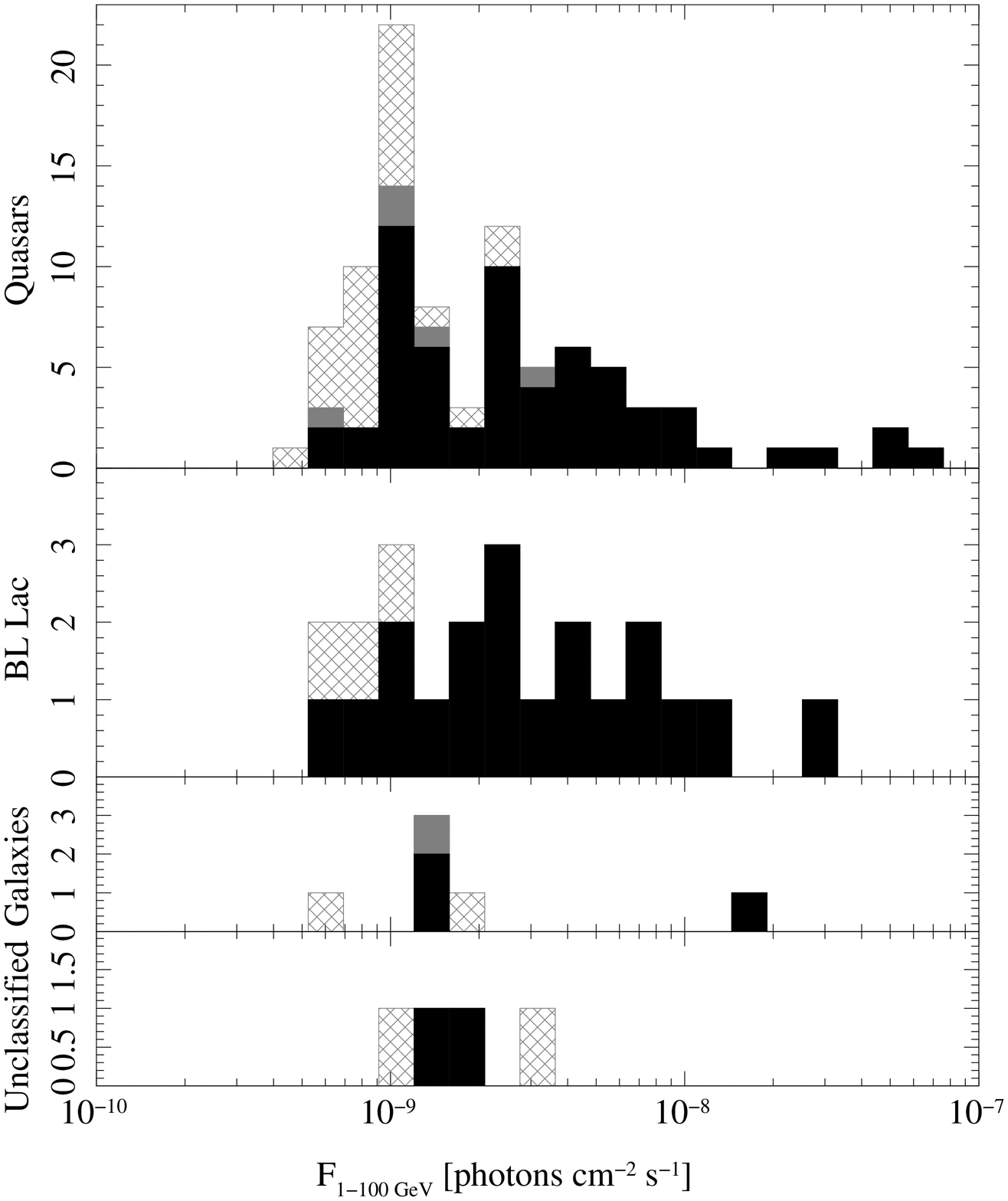}
      \hfill\includegraphics[width=\columnwidth]{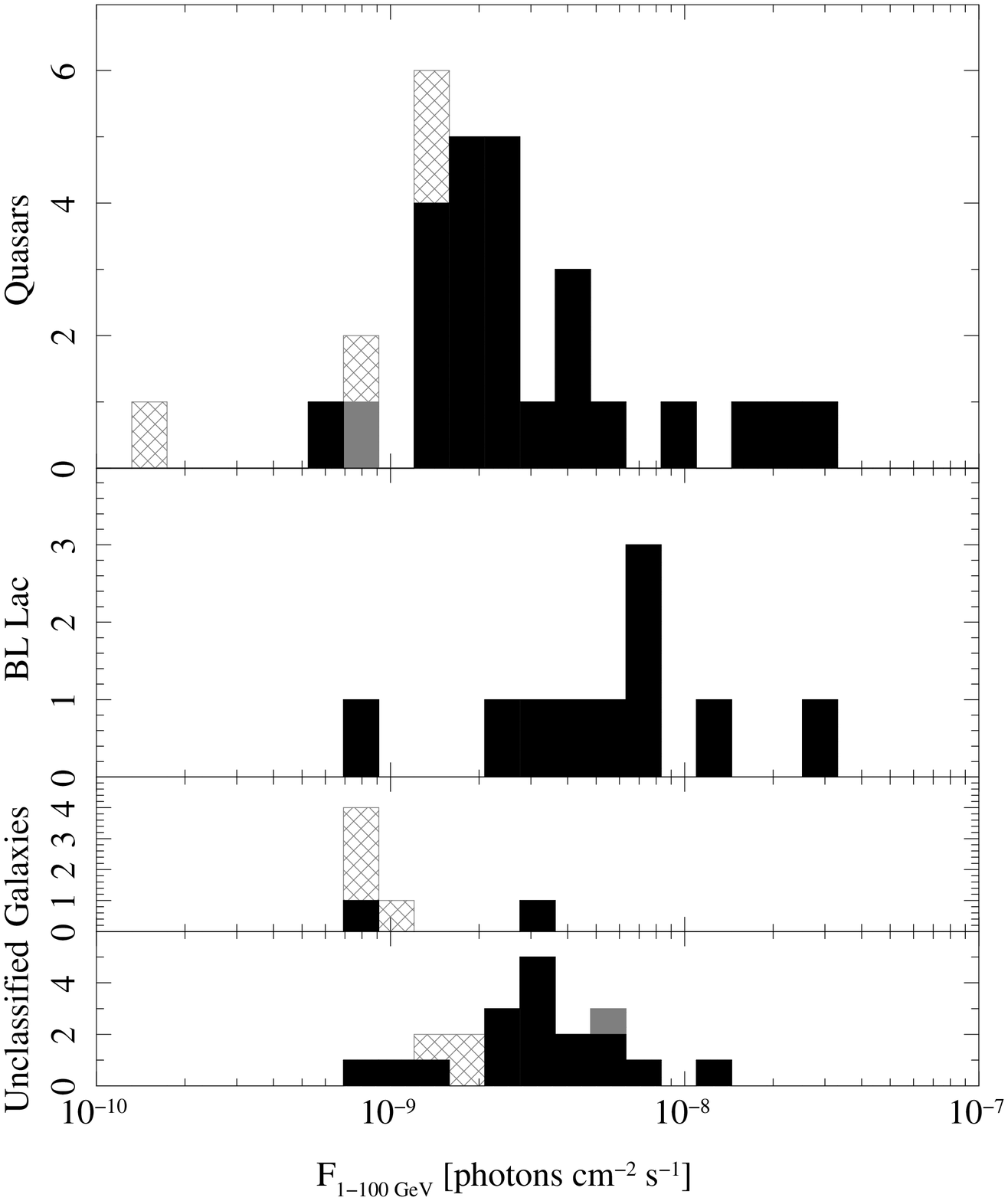}
      \caption{Distribution of the flux of the associated sources in the
      MOJAVE\,1 (left) and TANAMI (right) samples. Sources associated
      with catalog sources are shown in black, the tentative new
      detections in gray and the upper limits are shown cross hatched.}
      \label{fig:fluxdistr}
\end{figure*}

Figure~\ref{fig:flux_spec} indicates a difference between the MOJAVE\,1
and the TANAMI sample which is caused by the different selection
criteria. The $\gamma$-ray selected TANAMI sub-sample contains several
hard sources, i.e., sources with relatively small $\gamma$-ray photon
spectral index, including many BL\,Lac objects and unassociated sources.
BL\,Lac objects tend to have on average harder $\gamma$-ray spectra than
the quasars \citep{LBAS2009,1LAC}.

Hard sources can be better localized with \fermilat{}. Thus the limiting
photon flux for a $\textnormal{TS}>25$ detection is significantly lower
for BL\,Lac objects than for quasars with softer spectra
\citep[Fig.~10]{1LAC}. In Fig.~\ref{fig:fluxdistr} this effect is not
visible as the hardest sources in the sample do not reach a photon
spectral index of $\Gamma<1.5$ as it is the case for the sources
discussed by \citet{1LAC}. In addition, they consider the photon flux
above 100\,MeV instead of 0.1--100\,GeV, which increases the photon flux
for soft sources significantly.

\begin{figure*}
    \includegraphics[width=\columnwidth]{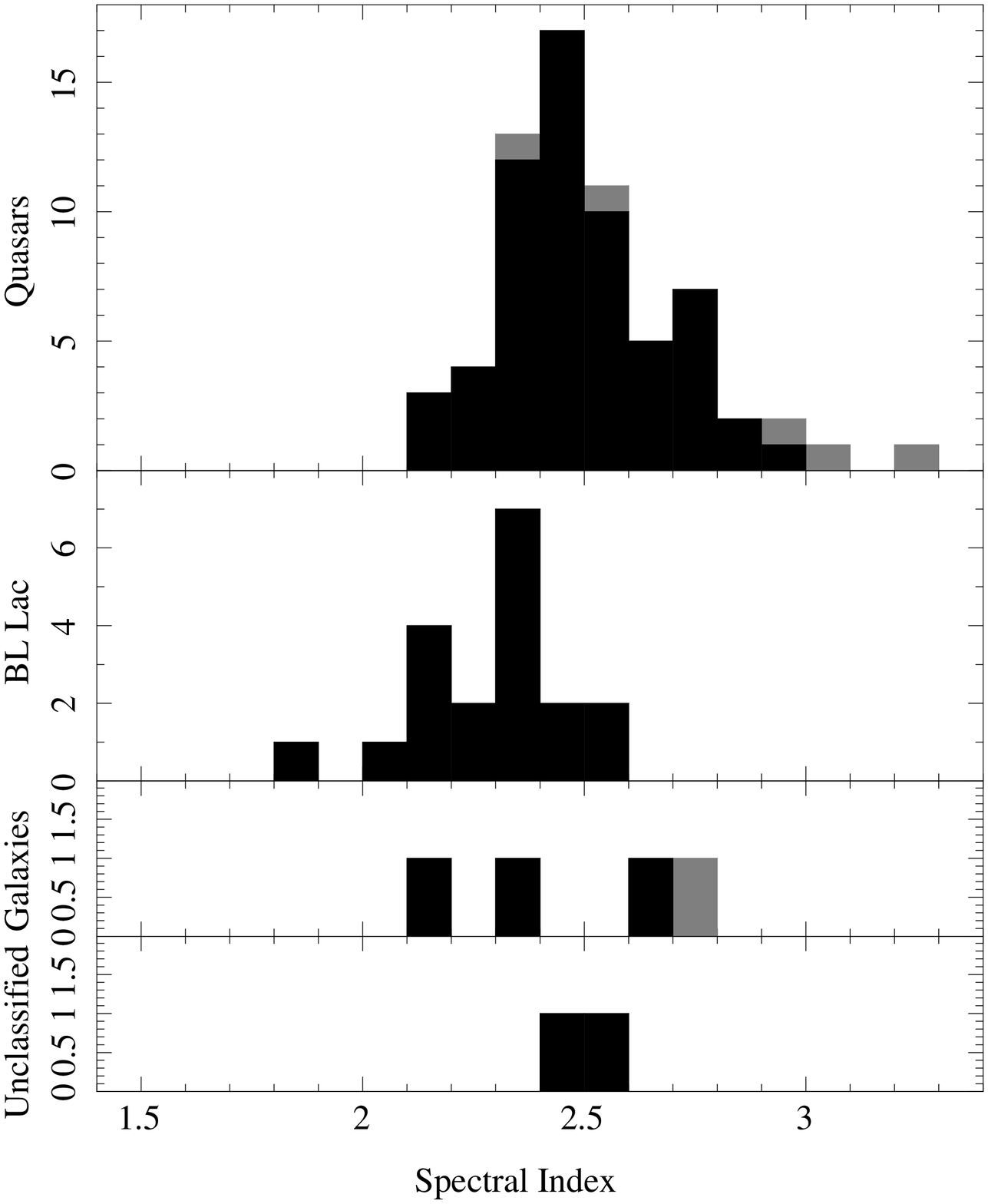}
      \hfill\includegraphics[width=\columnwidth]{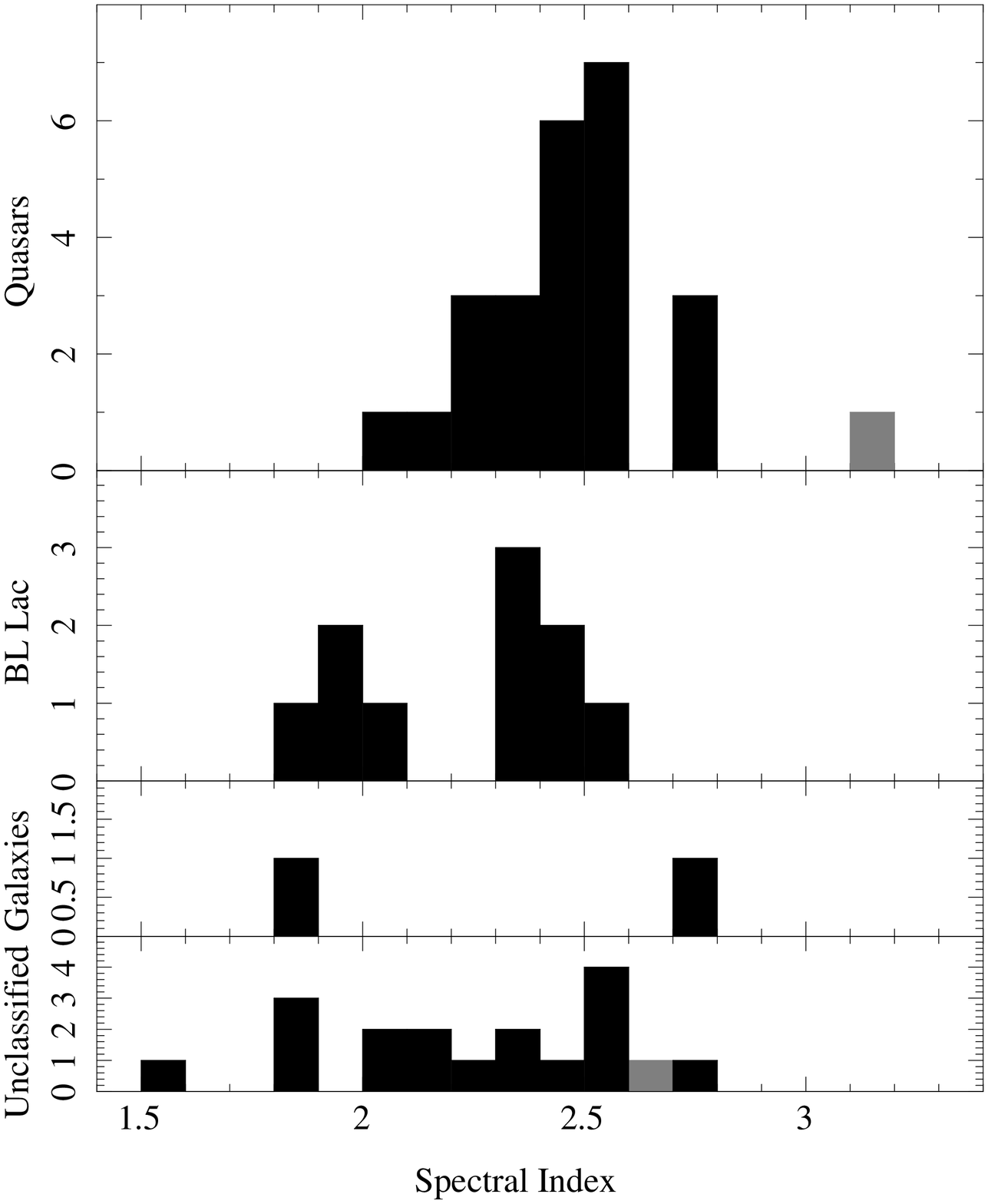}
      \caption{Distribution of the spectral index of the associated
      sources in the MOJAVE\,1 (left) and TANAMI (right) samples. The
      tentative new detections are shown in gray.} \label{fig:indxdistr}
\end{figure*}

The distribution of spectral indices (Fig.~\ref{fig:indxdistr}) confirms
that BL\,Lac objects exhibit harder $\gamma$-ray spectra than quasars
and shows that the majority of the spectral indices of the tentatively
new detected sources is softer than that of the associated sources,
which is consistent with the difficulty to detect soft weak sources with
\fermilat{}.

It is worth noting that all three BL\,Lac objects in the radio-selected
TANAMI sub-sample and the majority of BL\,Lac objects in the MOJAVE\,1
sample have $\gamma$-ray spectral indices $\Gamma > 2.2$. 
\citet{Abdo2010_LBAS_blazars} found that BL\,Lacs with $\Gamma > 2.2$
are mostly low synchrotron peaked (LSP) BL\,LACs \citep[see
also][Fig.~14]{1LAC}. It might be expected that a radio-selected sample
contains more LSP BL\,Lacs than high synchrotron peaked ones, as the LSP
type is brighter at the radio frequencies used for the selection.

\section{Summary}

The $\gamma$-ray properties of the MOJAVE\,1 and TANAMI AGN samples were
analyzed. The detection rates with \fermilat{} depend on the source
class. For the radio-selected MOJAVE\,1 sample and the radio-selected
TANAMI sub-sample BL\,Lac objects have the highest detection rates,
followed by the quasars, whereas only a small fraction of the radio
galaxies in the samples are seen by \fermilat{}. With a maximum
likelihood analysis on sources which were not associated with
$\gamma$-ray sources in the \fermilat{} First Source Catalog, we
obtained upper limits on the $\gamma$-ray flux of these sources and
found tentative new detections. 

The analysis of the upper limits and the tentative new detections will
be discussed in an upcoming paper.

\begin{acknowledgements}
    The \fermilat{} Collaboration acknowledges support from a number of
    agencies and institutes for both development and the operation of
    the LAT as well as scientific data analysis. These include NASA and
    DOE in the United States, CEA/Irfu and IN2P3/CNRS in France, ASI and
    INFN in Italy, MEXT, KEK, and JAXA in Japan, and the
    K.~A.~Wallenberg Foundation, the Swedish Research Council and the
    National Space Board in Sweden. Additional support from INAF in
    Italy and CNES in France for science analysis during the operations
    phase is also gratefully acknowledged. This research has been
    partially funded by the Bundesministerium f\"ur Wirtschaft und
    Technologie under Deutsches Zentrum f\"ur Luft- und Raumfahrt grant
    number 50OR0808. The Long Baseline Array is part of the Australia
    Telescope which is funded by the Commonwealth of Australia for
    operation as a National Facility managed by CSIRO.
\end{acknowledgements}

\end{document}